\documentclass{IEEEtran4PSCC}
\ifCLASSINFOpdf
   \usepackage[pdftex]{graphicx}
\else
   \usepackage[dvips]{graphicx}
\fi
\usepackage[cmex10]{amsmath}

\usepackage[dvipsnames]{xcolor} 

\usepackage{booktabs} 
\usepackage{multirow}  
\usepackage[section]{placeins} 

\makeatletter
\let\old@ps@headings\ps@headings
\let\old@ps@IEEEtitlepagestyle\ps@IEEEtitlepagestyle
\def\psccfooter#1{%
    \def\ps@headings{%
        \old@ps@headings%
        \def\@oddfoot{\strut\hfill#1\hfill\strut}%
        \def\@evenfoot{\strut\hfill#1\hfill\strut}%
    }%
    \def\ps@IEEEtitlepagestyle{%
        \old@ps@IEEEtitlepagestyle%
        \def\@oddfoot{\strut\hfill#1\hfill\strut}%
        \def\@evenfoot{\strut\hfill#1\hfill\strut}%
    }%
    \ps@headings%
}
\makeatother

\begin{document}
%
\title{Modelling Approaches of Power Systems Considering Grid-Connected Converters and Renewable Generation Dynamics}

\author{
\IEEEauthorblockN{Vinícius Albernaz Lacerda\\ Jaume Girona-Badia\\ Eduardo Prieto-Araujo\\Oriol Gomis-Bellmunt}
\IEEEauthorblockA{Centre d’Innovacio Tecnològica en Convertidors\\ Estatics i Accionaments (CITCEA-UPC) \\
Barcelona, Spain \vspace{-15pt}}
\and
\IEEEauthorblockN{Stephan Kusche\\ Florian Pöschke \\ Horst Schulte}
\IEEEauthorblockA{Department of Engineering I, Control Engineering \\
HTW Berlin - University of Applied Sciences\\
Berlin, Germany}
}


\maketitle


\begin{abstract}
This paper presents a comparative analysis of several modelling approaches of key elements used in simulations of power systems with renewable energy sources. Different models of synchronous generators, transmission lines, converters, wind generators and photovoltaic (PV) power plants are compared to assess the most suitable models for grid-connection studies. It also analyses how the dynamics of PV power plants and the mechanical dynamics of wind generators affect the electrical variables on the grid side. The models were compared in terms of precision and computational time through simulations of load connection, short-circuits, disconnection of generators and lines in a benchmark system modelled in Simulink. 
\end{abstract}

\begin{IEEEkeywords}
EMT simulation, mechanical dynamics, phasor simulation, power systems modelling, renewable generation dynamics. 
\end{IEEEkeywords}

\thanksto{\noindent This project has received funding from the European Union’s Horizon 2020 research and innovation programme under grant agreement No 883985 (POSYTYF project).}

\vspace{-8mm}
\section{Introduction}
The electrical power system is experiencing a deep penetration of renewable energy sources (RES) worldwide. Several countries have defined targets to increase the integration of RES, such as wind, solar, geothermal, hydro, ocean and biomass \cite{EC_2012,EC_2013}, using different solutions such as DC grids \cite{Van_Hertem_DCGrids}, microgrids \cite{microgrids_hatz} and Virtual Power Plants \cite{marinescu2021dynamic}.

In order to assess how present and future power systems will perform with high penetration of RES, researchers and industry need to use proper power systems models considering a variety of technologies. However, various modelling approaches have been proposed depending on the type of study, and there is not a single choice on how to model transmission lines, synchronous generators (SGs), converters and RES.

While relevant recommendations and guidelines were recently made available \cite{De_Carne_2019,Paolone_2020, Shah_2021}, those are based on the researchers' experience and do not perform systematic comparisons. Other systematic comparisons were also performed, but those focus on each component individually, such as power electronics \cite{Chiniforoosh_2010,Das_2016}, wind generation \cite{Ali_2011}, or simulation approaches \cite{Manzoni_2002, De_Rua_2020}, hence not analysing the interaction between different modelling approaches when key elements are simulated together. Thus, important questions about the suitability of each model in relation to other key elements of the system still need to be addressed.

Therefore, this paper presents a comparative analysis among several modelling approaches, considering different levels of detail of the main components of power systems with RES. It also considers the dynamics of photovoltaic (PV) power plants and mechanical dynamics of wind generators (WGs) and analyses how these dynamics affect the electrical variables on the grid side. 

The remainder of this paper is organized as follows. Section~\ref{sec_modelling_approaches} briefly introduces the models of SGs, transmission lines, converters and RES used in this study. Section~\ref{sec_methodology} presents the methodology of the comparative analysis, including the simulated system and the tests performed. The result are shown in Section~\ref{sec_results} followed by discussions. Finally, the conclusions are drawn in Section~\ref{sec_conclusions}.
\vspace{-2mm}
\section{Modelling approaches}\label{sec_modelling_approaches}
\vspace{-1mm}
SGs, transmission lines and converters are conventional elements used when simulating power systems with RES.
Several models of converters were proposed, varying from detailed models, in the semiconductors domain, to high-level phasor models \cite{Cigre_604}. Multiple choices can also be found for synchronous generators, transmission lines and wind generators. The proper choice for each element will depend on the level of detail, the phenomena being analysed and the time available for simulation. Next, the models analysed in this study are described. 
\vspace{-2mm}
\subsection{Wind turbine modelling}
\vspace{-1mm}
The wind energy system in this work is modeled as a variable-speed turbine equipped with a SG fed by a back-to-back converter that also establishes the connection to the electrical grid. The associated aeromechanical energy conversion is nonlinear and depends on the current wind speed and the turbine states such as rotor speed or pitch angle. Different control loops govern the operation of the wind energy system with the power control of the wind energy conversion system being dominant. Depending on the current operating point and thus wind speed, the power control either maximizes the power output in partial-load region, limits the power output to rated in full-load region or generates the desired power output following a setpoint signal \cite{Poeschke_2020}. For the multi-MW class, the resulting closed-loop dynamics have timescales in the range of seconds \cite{Bjork_2021,Poeschke_2021}, and thus may represent relevant dynamics for the interaction with other participating units in the electrical grid. 

To study the effects of including these dynamics within the power system simulation, two different models are used to display wind power. The first comprises algebraic power relations that statically portray the produced power depending on the current wind speed as only input to the model. Therein, the power produced by the wind turbine $P_w$ is calculated using the wind speed $v$ and a power coefficient look-up table for $c_P(\frac{\omega R}{v},\beta)$ depending on the rotor speed $\omega$ and the pitch angle $\beta$, such that the power is given by $P_w = \frac{1}{2}c_P(\frac{\omega R}{v},\beta)\rho\pi R^2 v^3$ \cite{Hansen_2015}, where $R$ denotes the rotor radius of the wind turbine. Whenever the power output setpoint is below the extractable power of the wind, the power output of the model is immediately set to the desired value. This modeling implies a perfect control and following of the turbine states. Thus, it neglects the dynamics when transitioning between operating points that are governed by the control loops and system characteristics such as rotor inertia or pitch dynamics.

\vspace{-1pt}
The second model is capable of displaying the interaction of the mechanical turbine states and the control loops to form a dynamical description. It relies on the Takagi-Sugeno modeling framework that uses a convex description of linear models to describe nonlinear dynamics. To derive the model, first a linearization of NREL's 5 MW reference turbine \cite{NREL5MW_2009} using FAST \cite{NREL_FAST_2005} is conducted at several operating point within the relevant operating space. Subsequently, an observer-based controller in the Takagi-Sugeno framework is designed with respect to the aforementioned power control problem using linear matrix inequalities to derive the necessary controller gains. Further, the pitch and generator torque dynamics are modelled as first-order transfer functions to account for limited pitch rates and the generator dynamics. Details about the applied modeling and control structure can be found in \cite{Poeschke_2020,Poeschke_2021}. The model including tower, blade and drivetrain dynamics is validated using FAST and provides a proper description of the aerodynamic conversion dynamics governed by the control loops. As a result, the model-based control design directly yields a closed-loop system description that can be implemented in power system simulations capable of portraying the nonlinear dynamics inherited in the wind energy conversion process. 
\vspace{-2mm}
\subsection{Photovoltaic power station modelling}
\vspace{-1mm}
In general, PV power station consists of multiple arrays of PV cells connected to the grid via an optional DCDC converter and an inverter. Though the DCDC converter is not present in all facilities, it is used in this work because it facilitates power tracking control.

The model of one PV cell is realised as explicit single diode model, i.e. a current source (photon current $i_{ph}$) in parallel with a diode (diode current $i_d$) in parallel with a resistor ($R_h$) to accommodate losses \cite{xiao2017}, such that
\begin{equation}
  i_{pv} = i_{ph} - 
	\underbrace{i_{s} \left[ \exp \left( 	\frac{q_e v_{pv}}{A_n k_B T_c} 	\right) - 1 \right]}_{i_d} 
	- \frac{v_{pv}}{R_{h}} .
\end{equation}
In this form, the diode I-V characteristics is described by the theory of Shockley \cite{shockley1949}, using the Boltzmann constant $k_B$, elementary charge $q_e$ and using the tunable parameters ideality factor of the diode $A_n$, photon current $i_{ph}$ and saturation current $i_s$. The tunable parameters are fitted on the I-V and P-V characteristics of the PV cell, obtained under standard test conditions (STC). The cell temperature $T_c$ is only subject to slow changes in time and therefore assumed to be constant at the STC value of 25$^\circ$C. On the other hand, the irradiation $S$ may change rapidly, and affects the photon current following ($\alpha_T$ being a PV cell dependent parameter):
\begin{equation}
	\frac{i_{ph}}{i_{ph}^{STC}} 
= \frac{S}{S^{STC}}\left[1 + \alpha_{T}(T_{c} - T_{c}^{STC})\right].
\end{equation}


Power transformation is done in a first step via a boost converter \cite{Erickson2007} and in a second step via an inverter. The boost converter is controlled via the duty cycle $D$, which determines the conversion ratio between PV voltage $v_{pv}$ and DC link voltage $v_{dc}$. Within the continuous conduction mode, $v_{pv} = (1-D) v_{dc}$ holds \cite{Erickson2001}. A higher level controller is used to track either the MPP or demand power point (DPP) using a perturb and observe (P\&O) method, e.g. \cite{Teulings1993,Kim1996,Femia2005}. 
Usually the feedback information is the PV power: $P_{pv}=v_{pv}\times i_{pv}$ (MPP). If we aim for a demanded power $P_{dem}$, feedback is replaced by $P^*_{pv}=-|P_{pv}-P_{dem}|$ (DPP). 
\vspace{-2mm}
\subsection{Synchronous machines modelling}
\vspace{-1mm}
\subsubsection{Simplified model}
\begin{figure}[htb]
	\vspace{-5pt}
	\centering
	\includegraphics[width=0.24\textwidth]{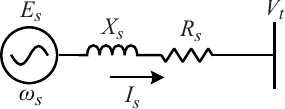}
	\caption{Simplified SG model single-line diagram.}
	\label{fig_simp_SG_diagram}
\end{figure}

The simplified SG model consists of a voltage-behind-impedance model with variable frequency, governed by the swing equation. The model diagram is depicted in Fig.~\ref{fig_simp_SG_diagram}, where mechanical speed $\omega_m$ and the internal voltage $E_s$ are calculated by
\begin{equation}
2H\frac{\mathrm{d} \omega_m}{\mathrm{d} t} = P_m-P_e
\end{equation}
\begin{equation}
E_s = \dfrac{1}{1 + \tau_f s} V_f
\end{equation}
where $H$ is the SG inertia constant, $P_m$ is the mechanical power, $P_e$ is the electrical power, $\tau_f$ is the field circuit time constant in p.u., and $V_f$ is the field voltage in p.u.

\vspace{-1mm}
The first-order transfer function with time constant $\tau_f$ links the SG internal voltage to the field voltage. Adding $\tau_f$ to the simplified model allows using the same excitation system used in detailed SG models. $\tau_f$ can be calculated from the SG's field resistance and inductance or numerically by fitting a step response in the field circuit of the complete model.  

\subsubsection{IEEE Model 2.2}
\begin{figure}[htb]
	\vspace{-7pt}
	\centering
	\includegraphics[width=0.48\textwidth]{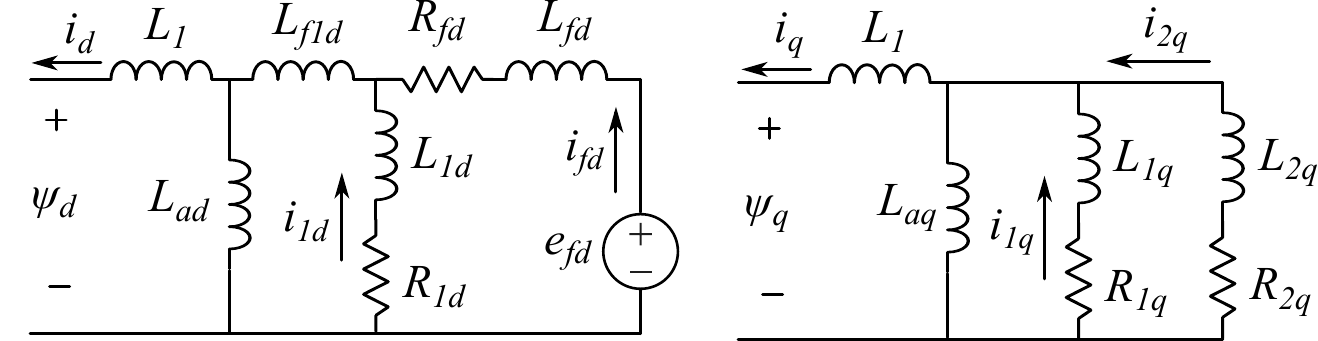}
		\vspace{-1mm}
	\caption{IEEE Model 2.2 single-line diagram.}
	\label{fig_SG22_diagram}
	\vspace{-3mm}
\end{figure}

The IEEE Model 2.2 \cite{IEEE_1110} is a precise yet simple electrical model in the $dq$ axis. This model takes into account the dynamics of the stator, field, and damper windings. One of the benefits is that standard data supplied by manufacturers is usually based on the inherited parameters \cite{IEEE_1110}. The equivalent circuit is represented in the rotor reference frame, depicted in Fig.~\ref{fig_SG22_diagram}, where the voltages are calculated as
\begin{subequations}
	\begin{align}
		v_d &= \frac{\mathrm{d} \psi_d }{\mathrm{d} t} - i_d R_s - \omega_s \psi_q \\
        v_q &= \frac{\mathrm{d} \psi_q }{\mathrm{d} t} - i_q R_s - \omega_s \psi_d 
	\end{align}
	\vspace{-10pt}
	\label{eq_SG_22}
\end{subequations} 

\subsubsection{IEEE 2.2 Model with saturation}
According to \cite{IEEE_1110}, the SG saturation significantly affects both rotor angle and excitation currents, thus it is important to investigate how this affects the whole system simulation. In this study, the saturation was modelled as factor $k$ added to the excitation, following the description and parameters of \cite{Van_Cutsem_UofL_2013,Van_Cutsem_IEEE_2020}.
    
\vspace{-2mm}
\subsection{Transmission lines modelling}
\vspace{-1mm}
Transmission lines play an important role in power system simulation, either by influencing steady-state power flow or by adding additional dynamics to the system. Three models were implemented in this study, described next.

    \subsubsection{PI model}
    The PI model is widely used in several power system studies due to its simplicity and suitability to model power flows and electromechanical transients. However, the PI model is only precise for a limited frequency range \cite{Watson_2018}. 
    
    \subsubsection{Bergeron model}
    The Bergeron model is also a constant frequency model, but it is more accurate than PI model because it considers the travelling waves trough the line. In this model there is no direct connection between the two line terminals. Voltages and currents at one end affect indirectly the other end after a time delay due to the travelling time.
    
    \subsubsection{Frequency-dependent model}
    The frequency-dependent model precisely represents the transmission line or cable throughout the whole frequency range of interest. The line resistances, inductances, capacitances and conductances are calculated from the physical line geometry and are frequency dependent. The travelling waves are accurately represented in this model and the travelling wave speed is also frequency-dependent, making it suitable to model electromagnetic transients \cite{Watson_2018}. 
    
	\vspace{-2mm}
\subsection{Converter modelling}
	\vspace{-1mm}
Several models have been proposed for the conventional two-level VSC, varying from very detailed models to high-level RMS models \cite{Cigre_604}. Two models were compared in this study, described next.

    \subsubsection{EMT average model}
    One widely-used EMT model of VSCs is the average (AVG) model. The AVG model neglects the converter's high-frequency switching and considers that the VSC electrical model on the AC side is simply a controlled voltage source, defined as an amplification of the modulation index \cite{Weixing_Lu_2003}.
    
    \subsubsection{Phasor model}
    Phasor models aim to capture only the slow dynamics of the power grids, such as the electromechanical transients, with time constants generally bigger than 100 ms. In this model, the grid differential equations are substituted by algebraic equations. As the phasors are assumed to be rotating at nominal angular speed, voltages and currents have their dynamics around 0 Hz instead of 50 or 60 Hz. This allows to dramatically increase the simulation time step and consequentially the simulation speed.
    
    In phasor simulation, the VSCs are represented as current sources with magnitude and angle defined by the control system. 
    As phasor simulation aims mainly to speed up and simplify simulations, several approximations can be performed in the VSC control system to allow bigger simulation time steps. In this study, the reference output current given by the outer loop is directly sent to the current source, thus the output current-loop dynamics are neglected.

\section{Methodology}\label{sec_methodology}
To assess the influence of the aforementioned models, several studies were performed using an adapted Cigre European HV transmission network benchmark system \cite{Cigre_575}, modelled in Simulink. The system is composed of four synchronous generators, eight transmission lines and one VSC. 

Table~\ref{tab_sim_models} summarizes the models simulated in each test.
   \begin{figure*}[t]
 	\vspace{-3pt}
 	\centering
	\includegraphics[width=0.70\textwidth]{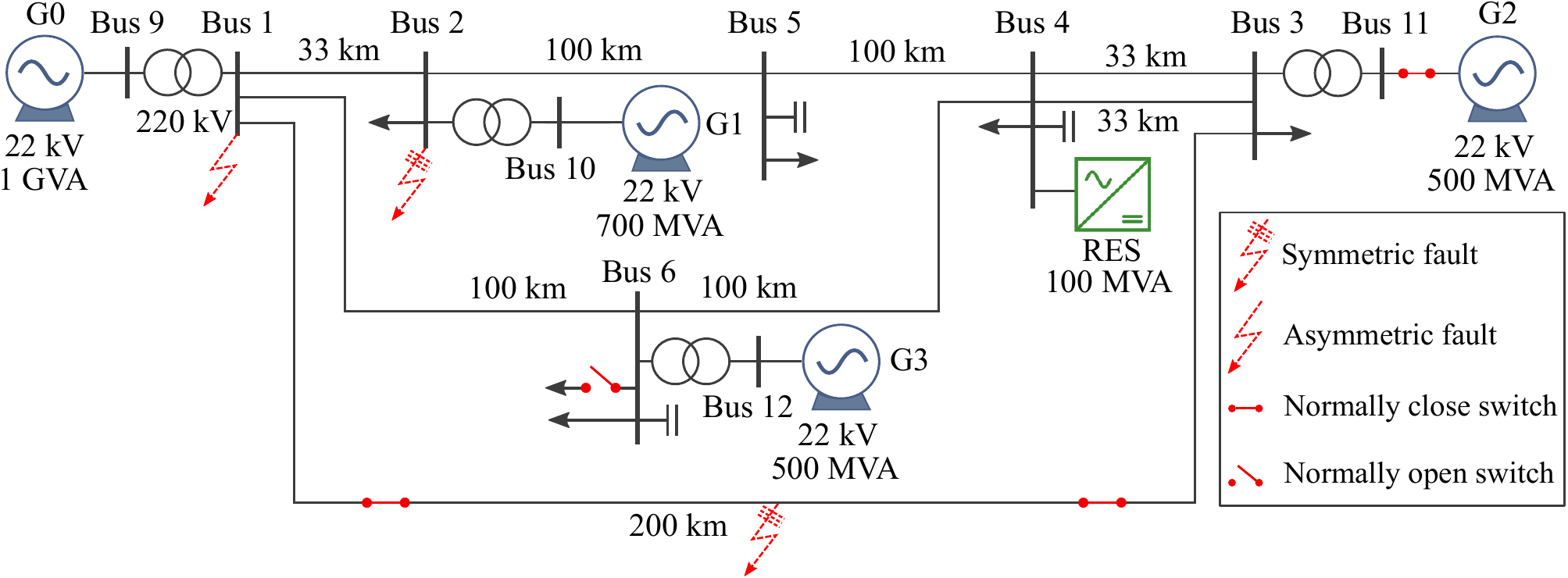}
 	\vspace{-3pt}
 	\caption{Simulated system single-line diagram. Modified from \cite{Cigre_575}.}
 	\label{fig_test_system_diagram}
 	\vspace{-7mm}
 \end{figure*}

    \subsubsection{Setpoint tracking}
    In this test, active and reactive power setpoints of the RES were set to 100~MW and 30~Mvar at $t \!=\! 1$~s and $t \!=\! 1~s$, respectively.

    \subsubsection{Load connection}
    In this test, a 100 MW, 20 Mvar load was connected to the bus 6 at $t \!=\! 5\,s$, dropping the system frequency and voltage.
    
    \subsubsection{Symmetrical faults}
    In this test, a 5~$\Omega$ three-phase fault was applied to bus 2 at $t \!=\! 5\,s$, lasting for 200 ms.
    
    \subsubsection{Asymmetrical faults}
    In this test, a 10~$\Omega$ mono-phase fault at phase B was applied to bus 1 at $t \!=\! 5\,s$, lasting for 500 ms.

    \subsubsection{Loss of generation}
    In this test, the generator G2 is disconnected from the system at $t \!=\! 5\,s$, producing a slow but large transient in the system.

    \subsubsection{Line outage}
    In this test, a permanent 1~$\Omega$ three-phase fault was applied to the line connecting bus 1 to bus 3. The line was isolated by two ideal circuit breakers 100~ms after the fault.
    
All tests were simulated using a fixed time step and the Euler method (\textit{ode1} in Simulink). The single-line diagram of the simulated system is depicted in  Fig.~\ref{fig_test_system_diagram}. 
\vspace{-4mm}
\begin{table}[h]
\caption{Simulated models}
\label{tab_sim_models}
\begin{tabular}{@{}ll@{}}
\toprule
\multicolumn{1}{c}{Component} & \multicolumn{1}{c}{Model} \\ \midrule
\multirow{3}{*}{Synchronous generator} & 
\multirow{3}{*}{\begin{tabular}[c]{@{}l@{}}Model 2.2 \cite{IEEE_1110} with saturation \\ \textbf{model 2.2 \cite{IEEE_1110} without saturation}\\ inertia-only model\end{tabular}} \\
 &  \\[12pt]
\multirow{3}{*}{Transmission line} & \multirow{3}{*}{\begin{tabular}[c]{@{}l@{}}Frequency-dependent (FD) model \\ Bergeron model,\\ \textbf{nominal PI model}\end{tabular}} \\
 &  \\[12pt]
Converter & \textbf{EMT average model}, Phasor model \\[3pt]
\multirow{3}{*}{Renewable generation} & \multirow{3}{*}{\begin{tabular}[c]{@{}l@{}} Dynamic PV power plant, dynamic wind turbine, \\ algebraic power laws, \\\textbf{ideal DC voltage source} \end{tabular}}\\
\\
 &  \\ \bottomrule
\end{tabular}
\vspace{-4mm}
\end{table}

The tests were performed as follows. First, one model of each element (SG, transmission lines, converter, RES) was chosen to form a base group, which are indicated in bold in Table~\ref{tab_sim_models}. Afterwards, each test was performed varying one element per time in relation to the base group. This allowed to identify the influence of each model in the overall system behaviour.

Three key aspects were analysed for each model: Precision and execution time. Finally, the key aspects were combined to express the suitability of each model for the evaluated scenarios. 




\vspace{-3mm}
\section{Results and Discussion}\label{sec_results}

\vspace{-1mm}This section presents the simulation results, followed by discussions.
Due to the extensive number of tests, only a few cases and variables are presented, which are representative of each test. 
	\vspace{-2mm}
\subsection{Influence of synchronous generator model}
\begin{figure}[htb]
	\vspace{-10pt}
	\centering
	\includegraphics[width=0.48\textwidth]{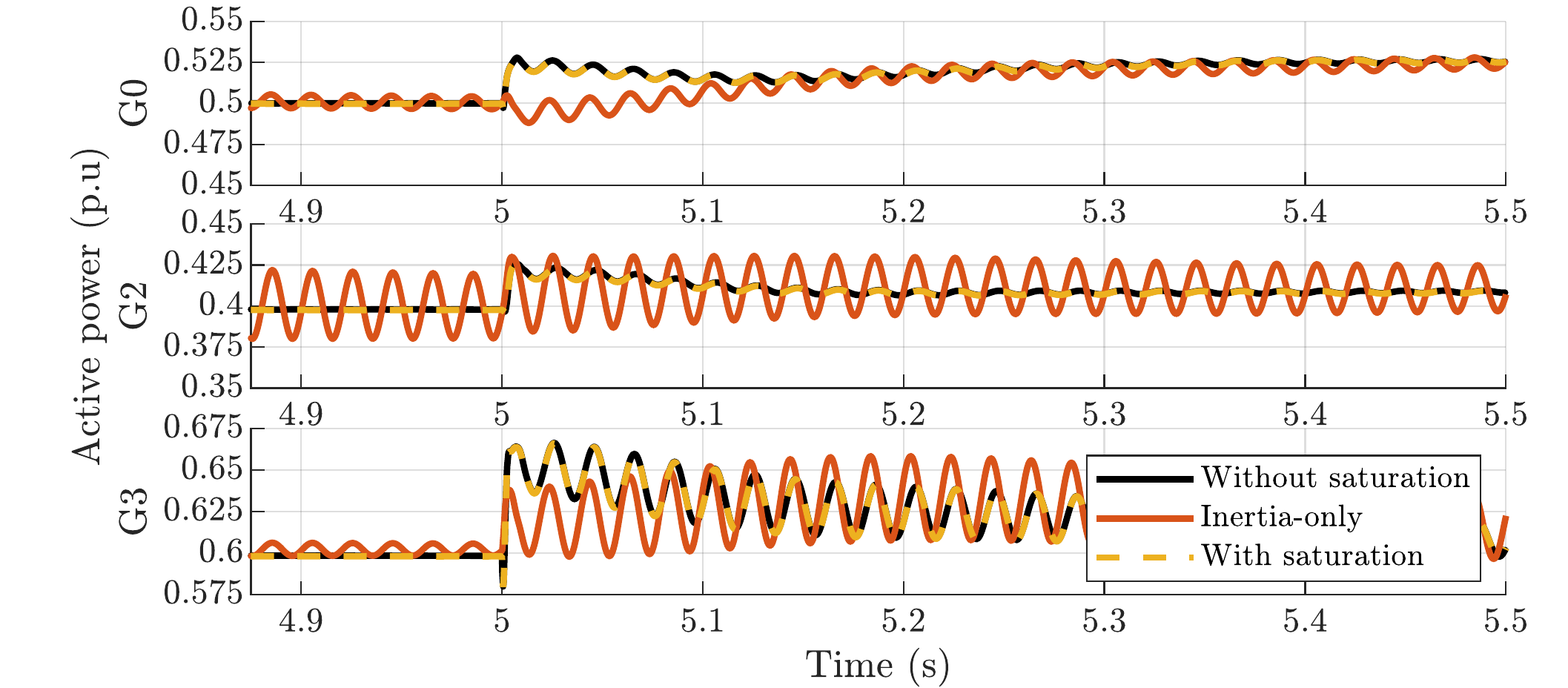}
	\vspace{-3mm}
	\caption{SGs active power during a load connection for each SG model.}
	\label{fig_gen_connection_power_gen}
	\vspace{-8mm}
\end{figure}

\begin{figure}[htb]
	\vspace{-20pt}
	\centering
	\includegraphics[width=0.48\textwidth]{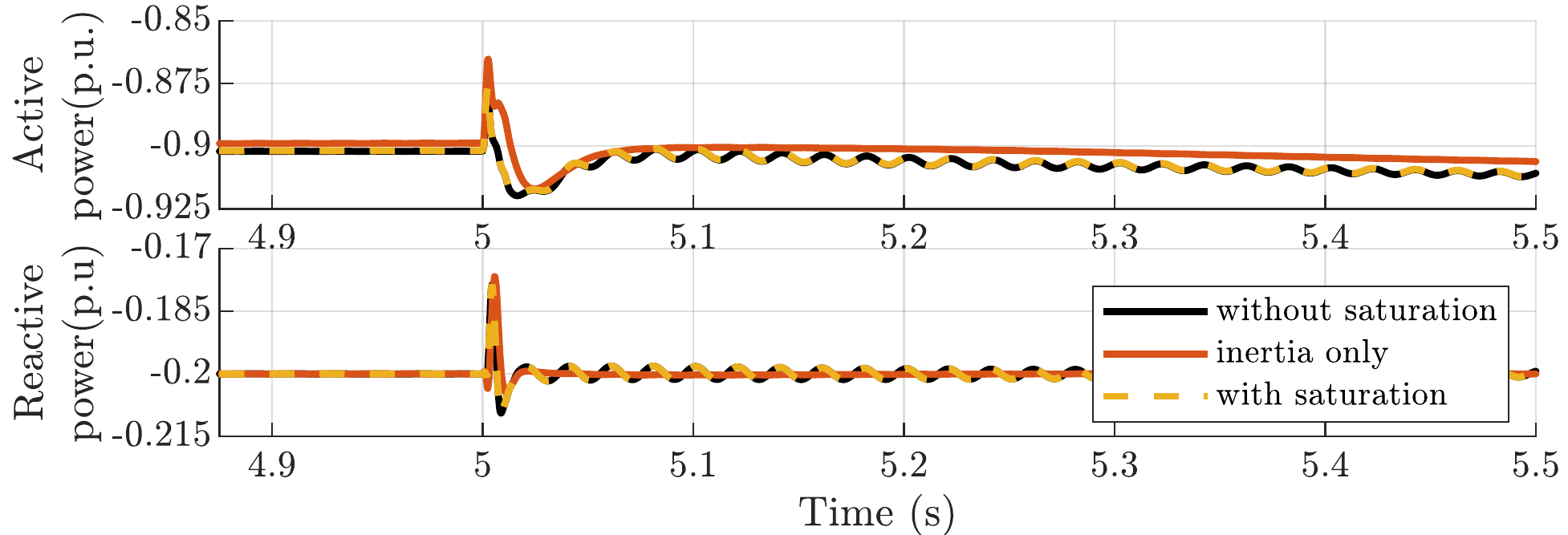}
	\vspace{-3mm}
	\caption{RES power during a load connection for each SG model.}
	\label{fig_gen_connection_power_VSC}
	\vspace{-2mm}
\end{figure}

\begin{figure}[htb]
	\vspace{-10pt}
	\centering
	\includegraphics[width=0.48\textwidth]{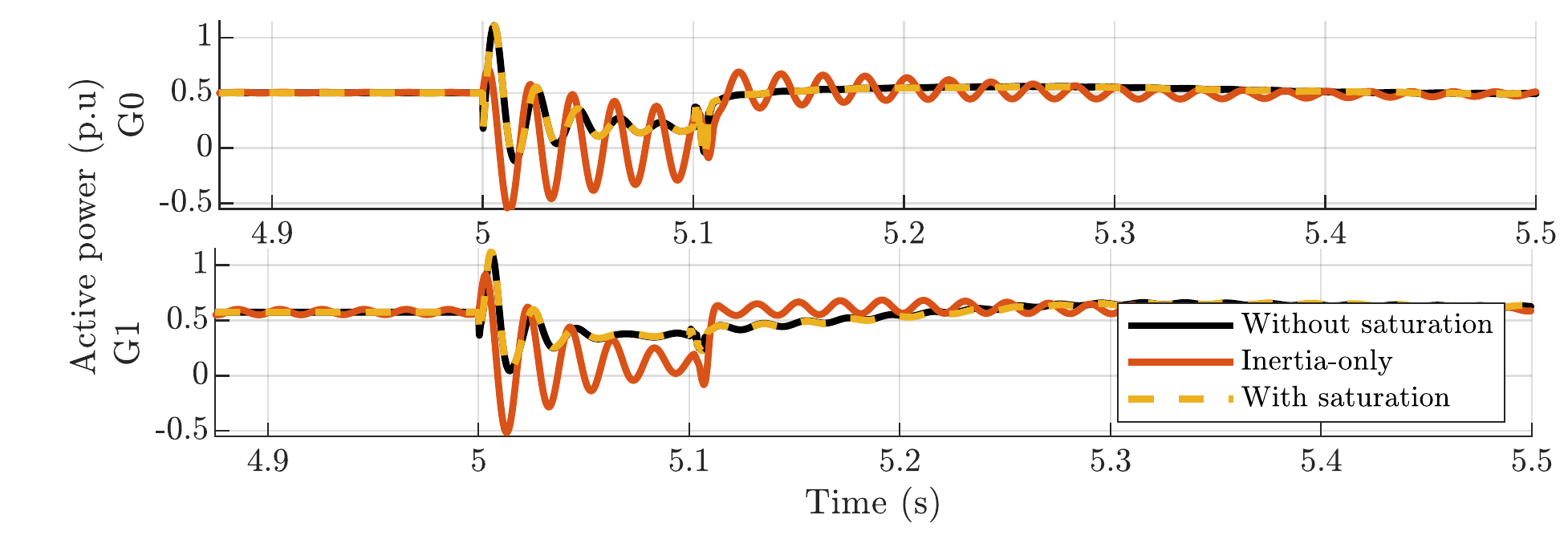}
	\vspace{-3mm}
	\caption{SGs active power during a line outage for each SG model.}
	\label{fig_gen_line_power_gen}
	\vspace{-1mm}
	
\end{figure}

\begin{figure}[htb]
	\vspace{-10pt}
	\centering
	\includegraphics[width=0.48\textwidth]{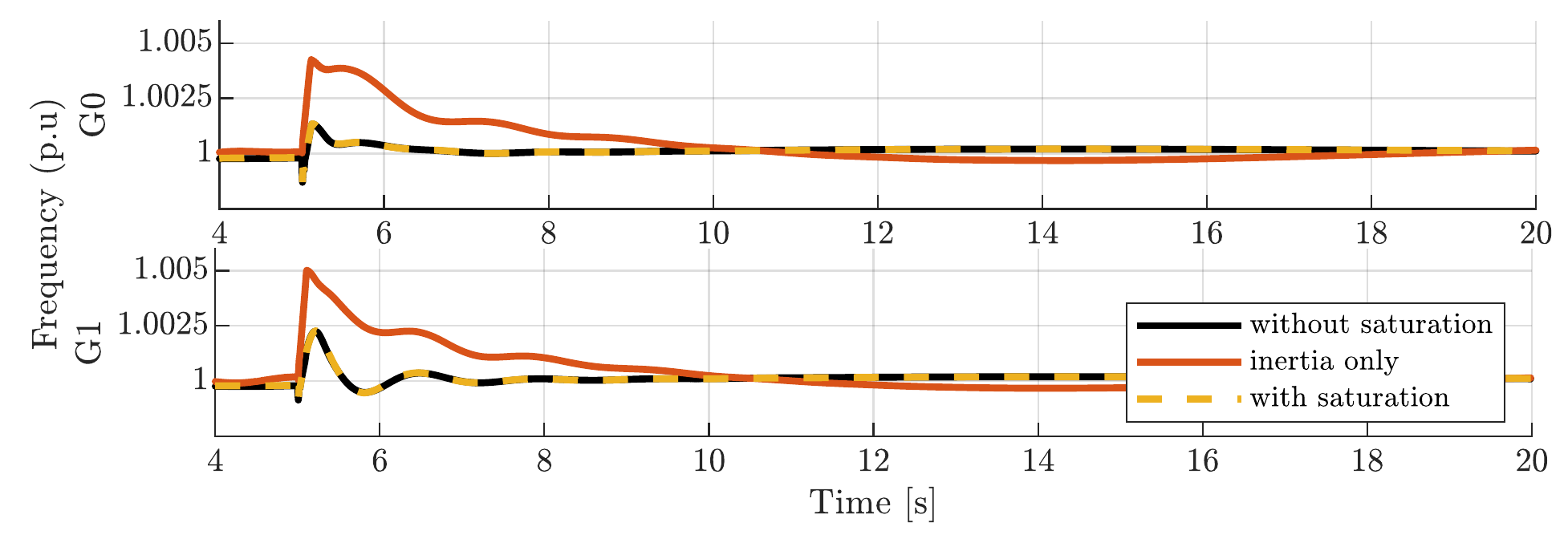}
	\vspace{-3mm}
	\caption{SGs speed during a line outage for each SG model.}
	\label{fig_gen_line_freq_gen}

\end{figure}

\begin{figure}[htb]
	\vspace{-12pt}
	\centering
	\includegraphics[width=0.48\textwidth]{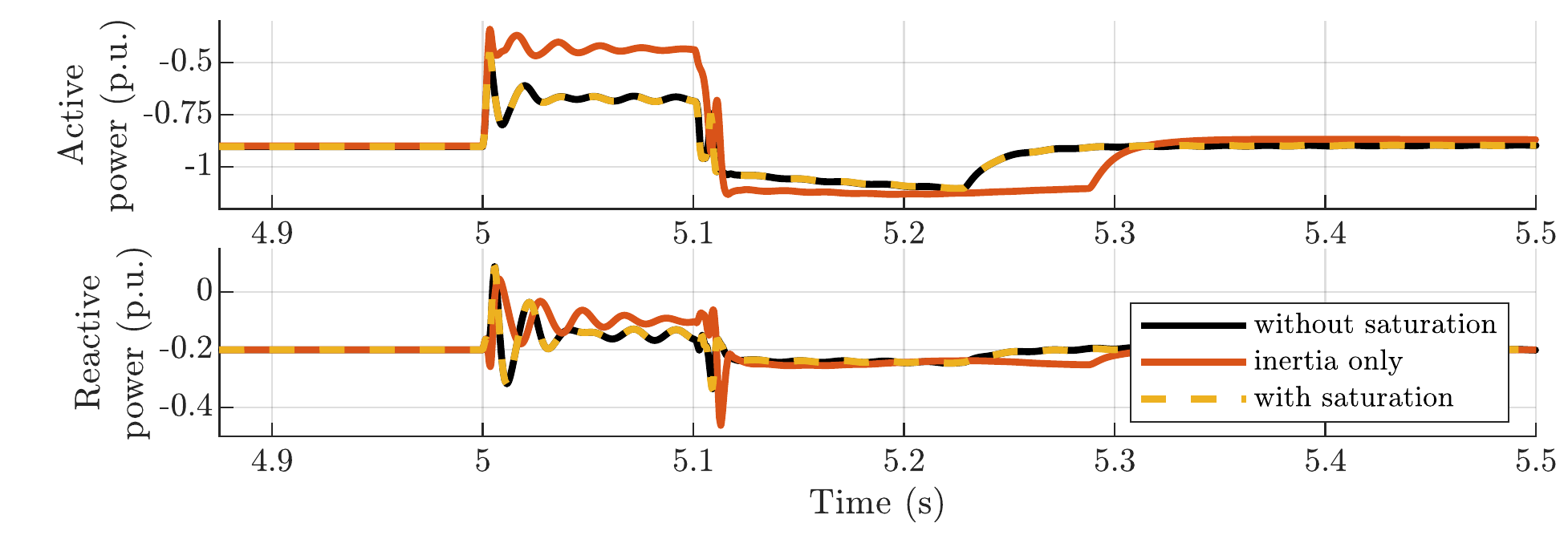}
	\vspace{-3mm}
	\caption{RES power during a line outage for each SG model.}
	\label{fig_gen_line_power_VSC}
	\vspace{-6mm}
\end{figure}

	\vspace{-1mm}
Figures~\ref{fig_gen_connection_power_gen}-\ref{fig_gen_line_power_VSC} present active power and speed for tests 1 and 6, respectively. From Figs.~\ref{fig_gen_connection_power_gen}-\ref{fig_gen_line_power_VSC} it can be observed that SGs' model 2.2 with and without saturation show a similar response. However, the system was simulated with light load, which can present optimistic results. When heavy load is simulated with contingencies, saturation should be considered. The inertia-only model, although requiring less parameters to simulate, presented more severe voltage oscillations due to the lack of damper windings. This effect is more noticeable during and post fault, as shown in Fig.~\ref{fig_gen_line_power_gen}. 

The SGs' electric model slightly influenced the system's total simulation time. Model 2.2 without saturation was 12 $\%$ faster than the inertia-only model and 18.4$\%$ faster than the model 2.2 with saturation.
\begin{figure}[htb]
	\vspace{-2pt}
	\centering
	\includegraphics[width=0.48\textwidth]{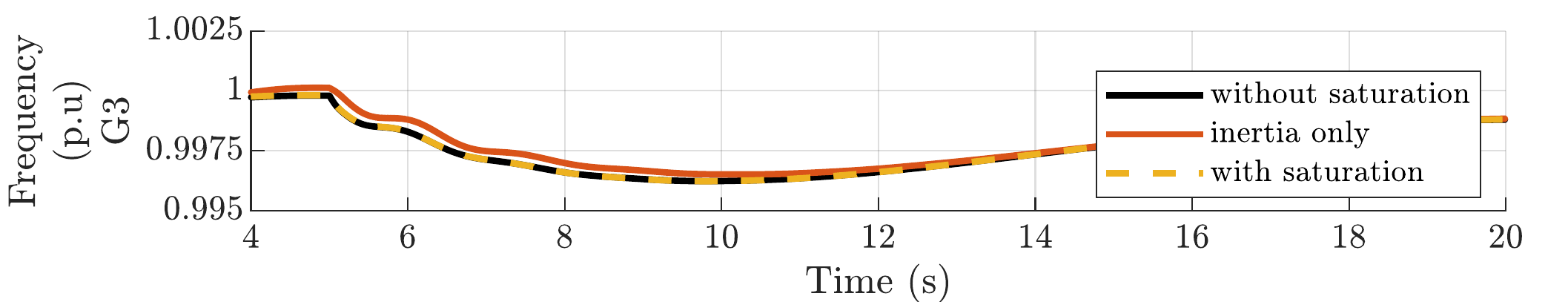}
	\vspace{-3mm}
	\caption{SGs speed during a load connection for each SG model.}
	\label{fig_gen_connection_freq_gen}
	\vspace{-1mm}
\end{figure}

\subsection{Influence of the transmission lines model}
\begin{figure}[htb]
	\vspace{-20pt}
	\centering
	\includegraphics[width=0.48\textwidth]{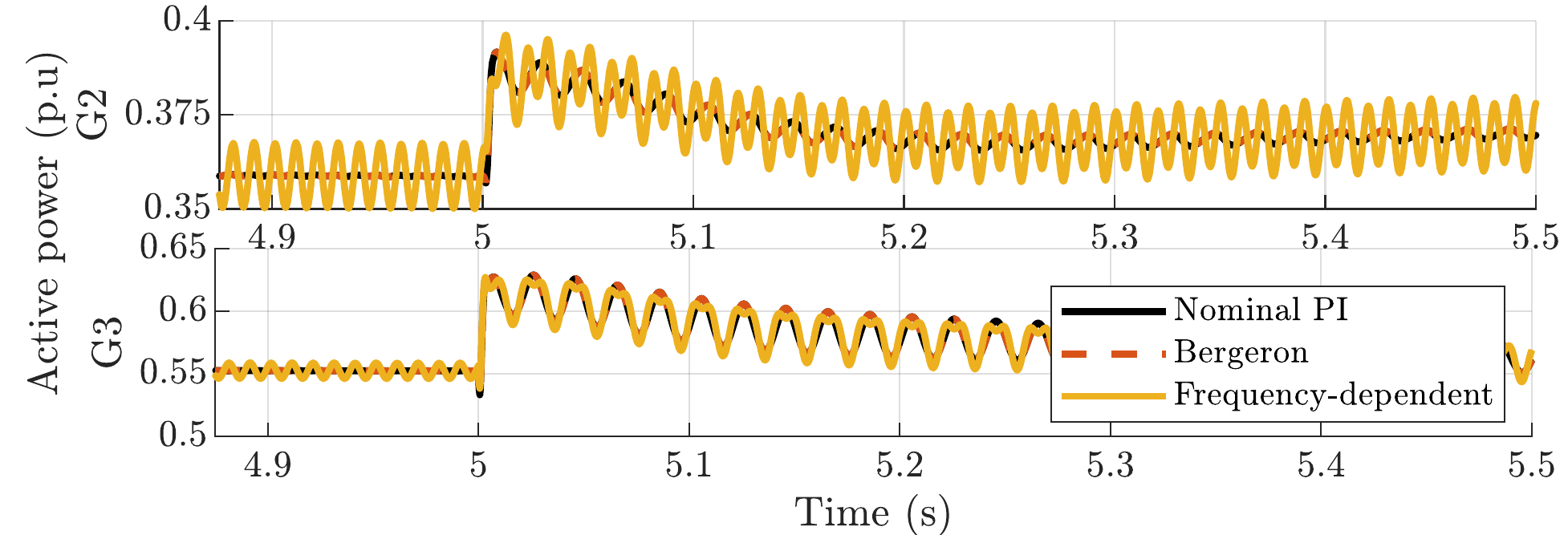}
	\vspace{-3mm}
	\caption{SGs active power during a load connection for each line model.}
	\label{fig_lines_load_power_gen}
\end{figure}

\begin{figure}[htb]
	\vspace{-10pt}
	\centering
	\includegraphics[width=0.48\textwidth]{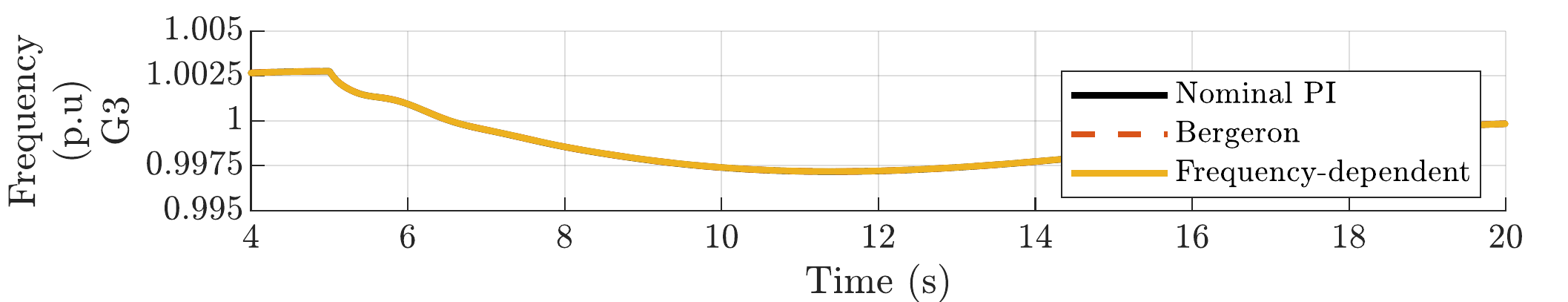}
	\vspace{-3mm}
	\caption{SGs speed during a load connection for each line model.}
	\label{fig_lines_load_freq_gen}
	\vspace{-4mm}
\end{figure}

\begin{figure}[htb]
	\vspace{-10pt}
	\centering
	\includegraphics[width=0.48\textwidth]{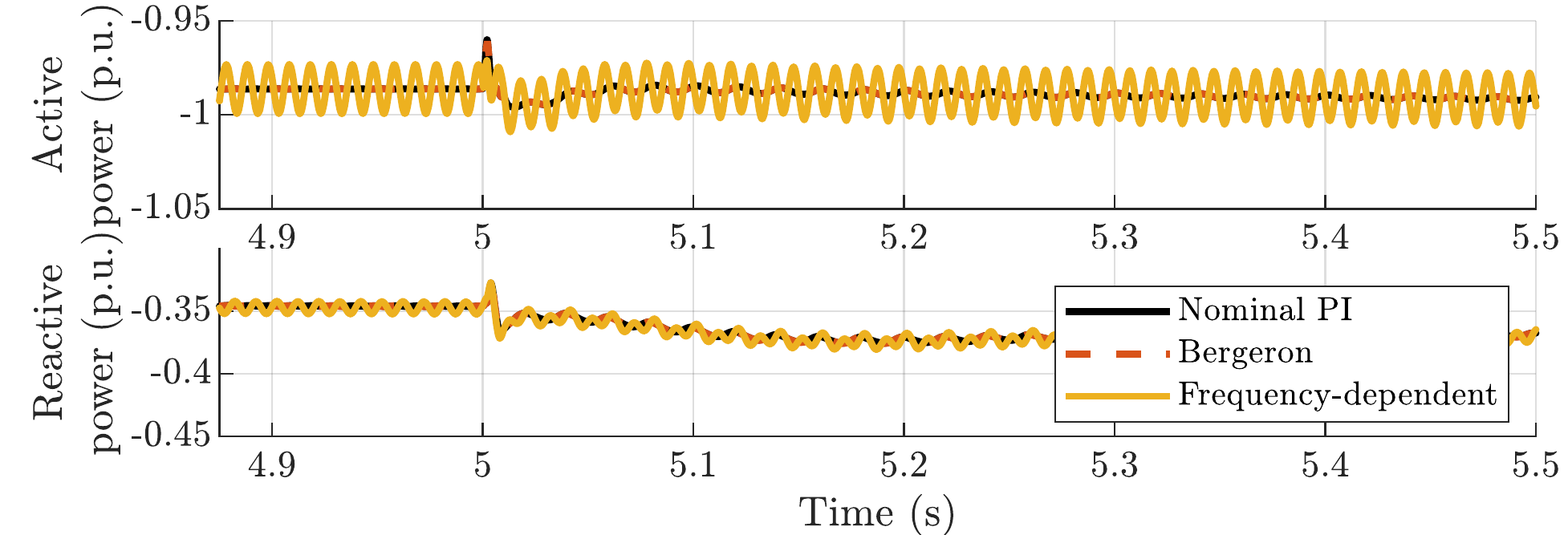}
	\vspace{-3mm}
	\caption{RES power during a load connection for each line model.}
	\label{fig_lines_load_power_VSC}
\end{figure}

\begin{figure}[htb!]
	\vspace{-10pt}
	\centering
	\includegraphics[width=0.48\textwidth]{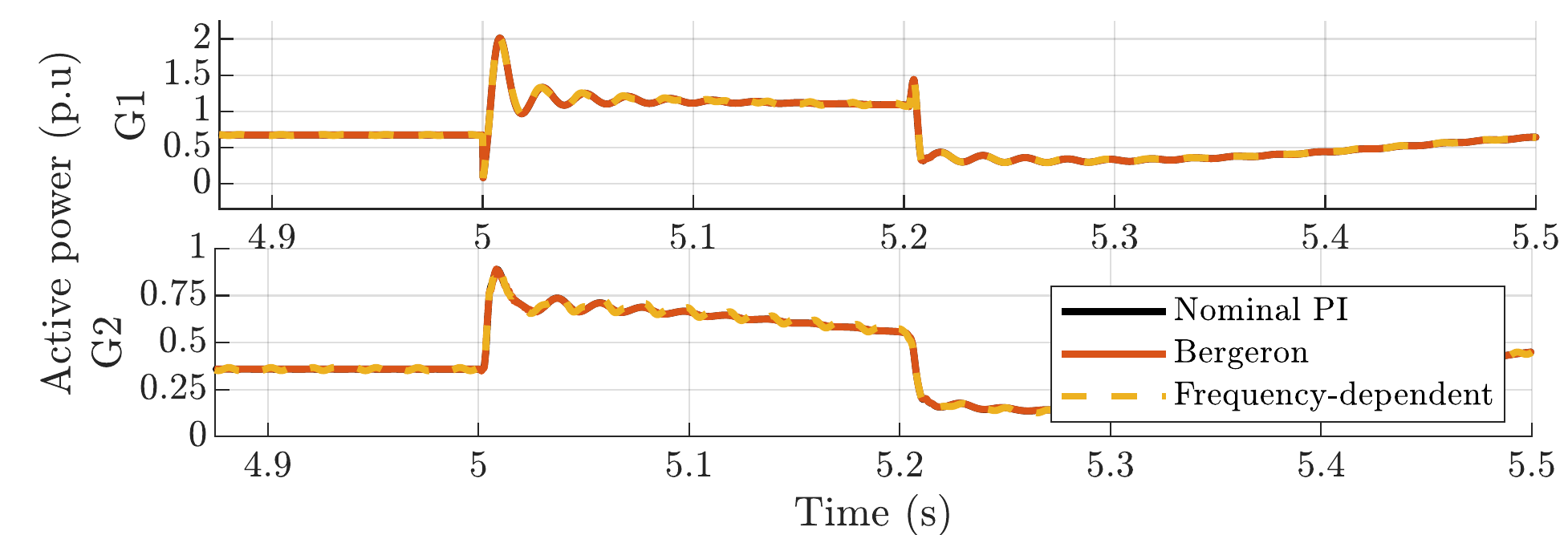}
	\vspace{-3mm}
	\caption{SGs active power during a symmetric fault for each line model.}
	\label{fig_lines_symetric_power_gen}
\end{figure}

\begin{figure}[htb!]
	\vspace{-10pt}
	\centering
	\includegraphics[width=0.48\textwidth]{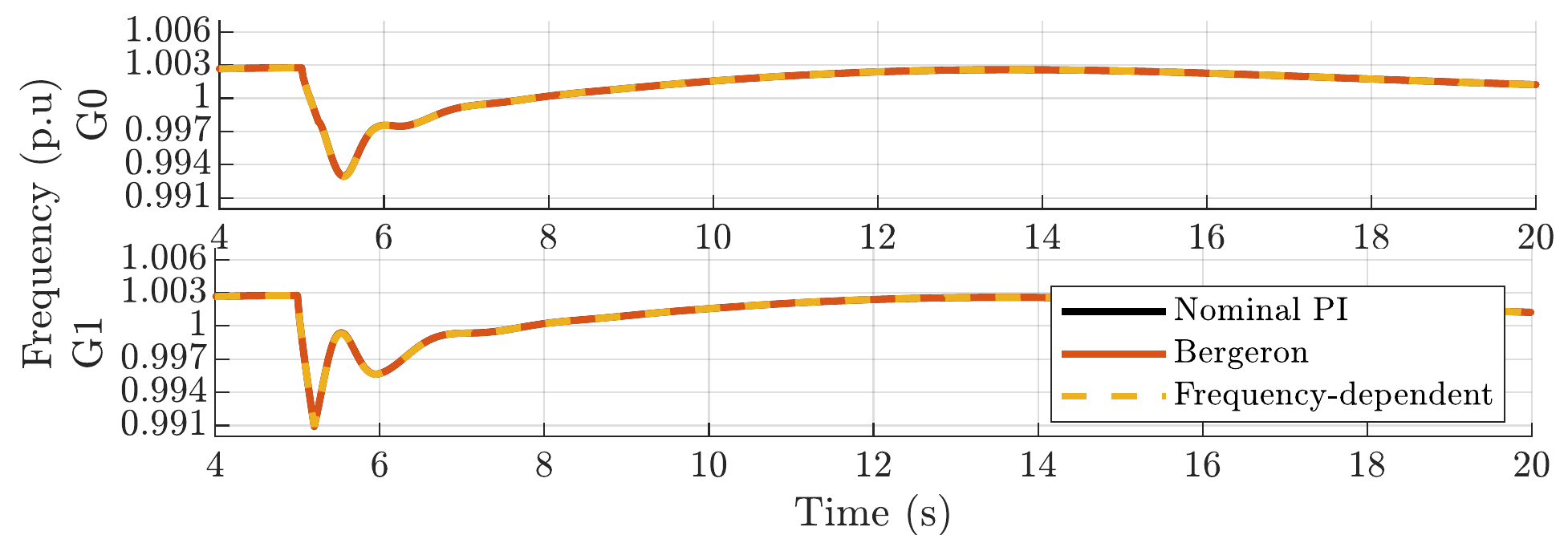}
	\vspace{-3mm}
	\caption{SGs speed during a symmetric fault for each line model.}
	\label{fig_lines_symetric_freq_gen}
	\vspace{-2mm}
\end{figure}

\begin{figure}[htb!]
	\vspace{-5pt}
	\centering
	\includegraphics[width=0.48\textwidth]{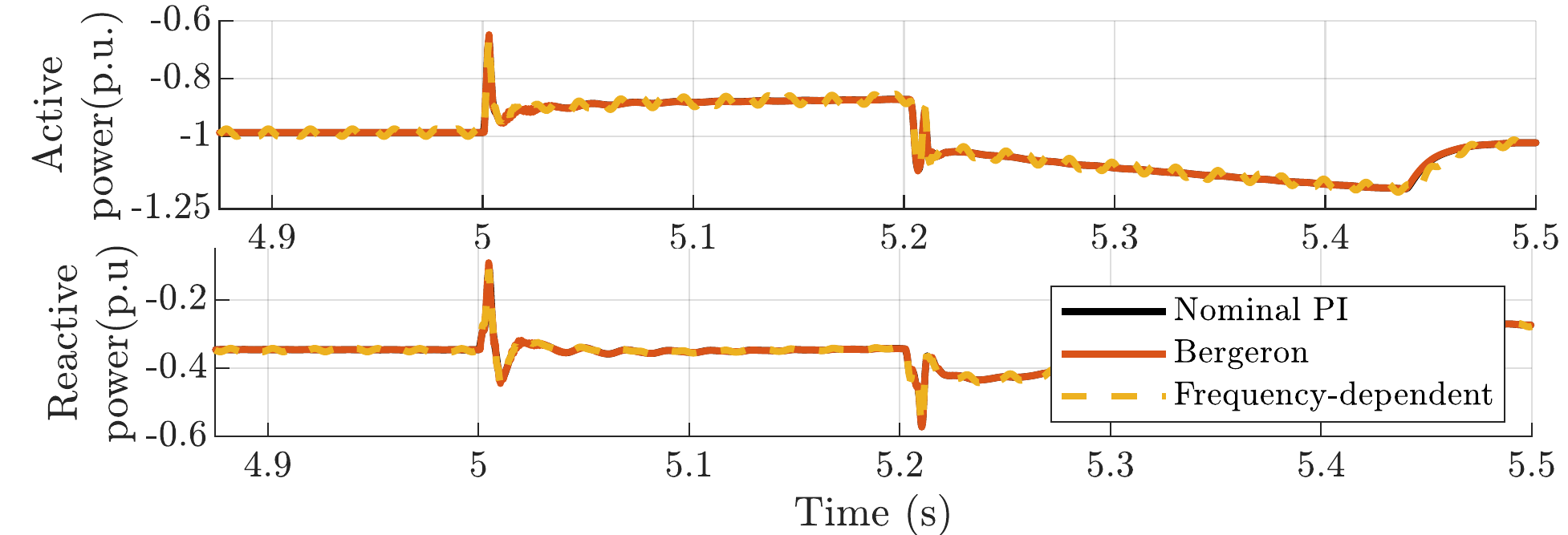}
	\vspace{-3mm}
	\caption{RES power during a symmetric fault for each line model.}
	\label{fig_lines_symetric_power_VSC}
	\vspace{-3mm}
\end{figure}

Figures \ref{fig_lines_load_power_gen}-\ref{fig_lines_load_power_VSC} and \ref{fig_lines_symetric_power_gen}-\ref{fig_lines_symetric_power_VSC} show the results for tests 2 and 3 (load connection and symmetric fault), respectively.  

In can be observed in both tests that the FD model represents higher frequencies due to wave propagation and resonance frequencies. These high-frequency components are also present though less represented in the Bergeron model and almost absent in PI nominal. However, the same low-frequency tendency is well represented in the three models. 
The symmetric fault. The fastest simulation was performed with the PI model, followed by the Bergeron model 16$\%$ slower on average than the the simulation with PI model. The FD model resulted in simulations 2113.6$\%$ slower than the PI model on average. 

Therefore, if the studies to be performed are essentially electromechanical transients, the PI model would be adequate. Using high-order transmission lines would add minor precision at the cost of a larger simulation time. Nevertheless, if electromagnetic transients are simulated or if the converter switching or control dynamics are a concern, the PI model might under-represent important high-frequency components, yielding optimistic results. The Bergeron model might be adequate between both extremes, where the nominal frequency dominates the response but the simulated lines' length are too large to neglect travelling wave effects.

\subsection{Influence of the converter model}
\begin{figure}[htb!]
	\vspace{-1mm}
	\centering
	\includegraphics[width=0.48\textwidth]{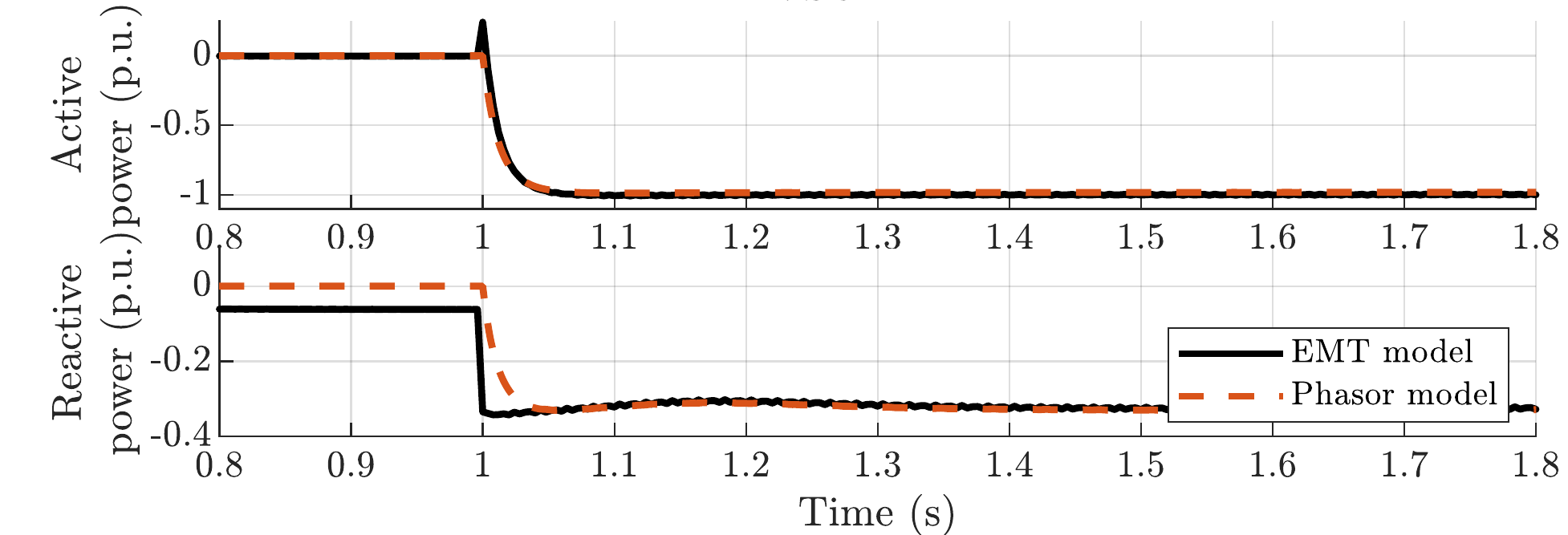}
	\vspace{-3mm}
	\caption{RES power during a setpoint tracking for each VSC model.}
	\label{fig_vsc_setpoint_power_VSC}
	\vspace{-1mm}
\end{figure}

\begin{figure}[htb!]
	\vspace{-10pt}
	\centering
	\includegraphics[width=0.48\textwidth]{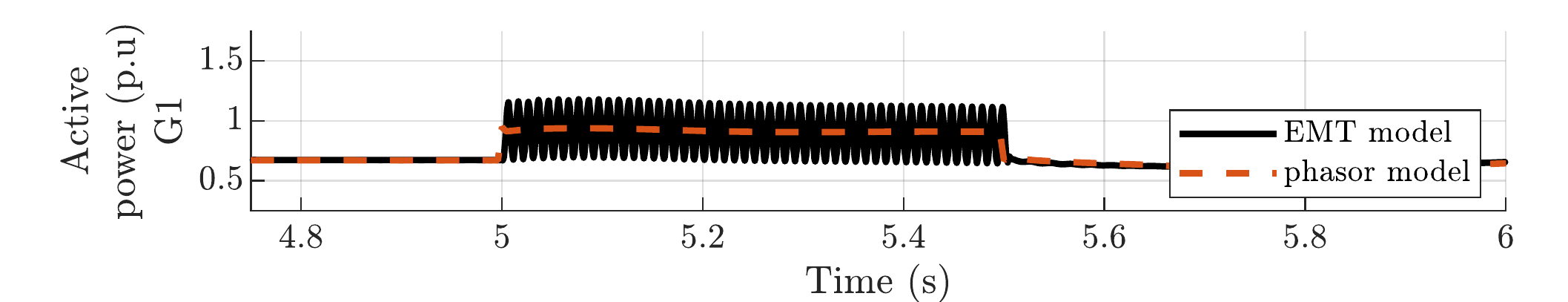}
	\vspace{-3mm}
	\caption{SG active power during an asymmetric fault for each VSC model.}
	\label{fig_vsc_asymetric_power_gen}
	\vspace{-8mm}
\end{figure}

\begin{figure}[htb]
	\vspace{-10pt}
	\centering
	\includegraphics[width=0.48\textwidth]{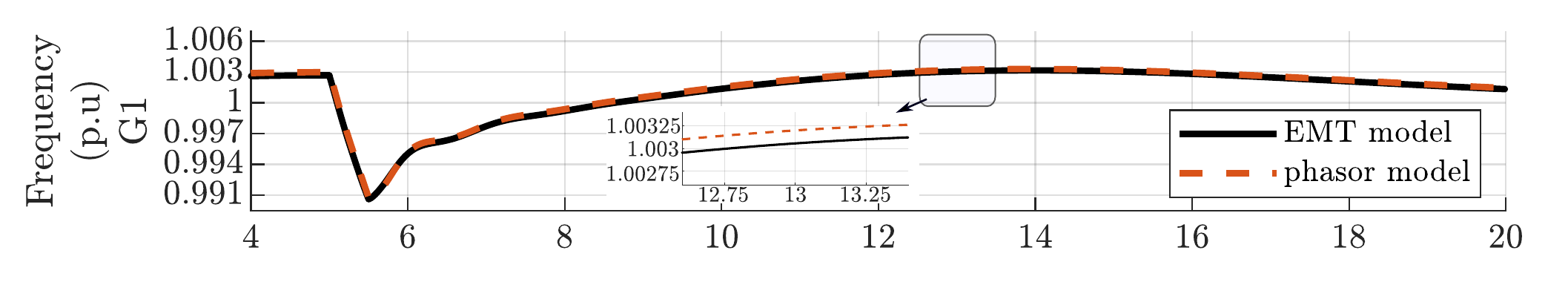}
	\vspace{-3mm}
	\caption{SG speed during an asymmetric fault for each VSC model.}
	\label{fig_vsc_asymetric_freq_gen}
	
\end{figure}

\begin{figure}[htb]
	\vspace{-13pt}
	\centering
	\includegraphics[width=0.48\textwidth]{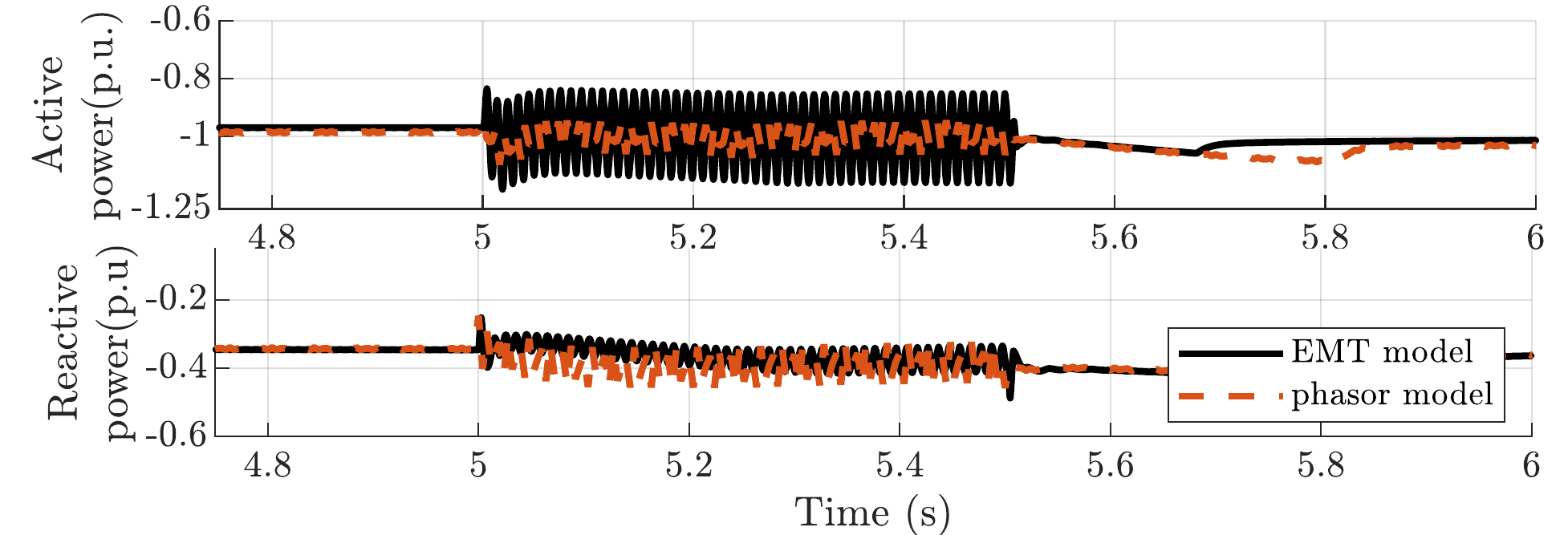}
	\vspace{-3mm}
	\caption{RES power during an asymmetric fault for each VSC model.}
	\label{fig_vsc_asymetric_power_VSC}
	\vspace{-7mm}
\end{figure}

\begin{figure}[htb]
 	\vspace{-20pt}
 	\centering
 	\includegraphics[width=0.48\textwidth]{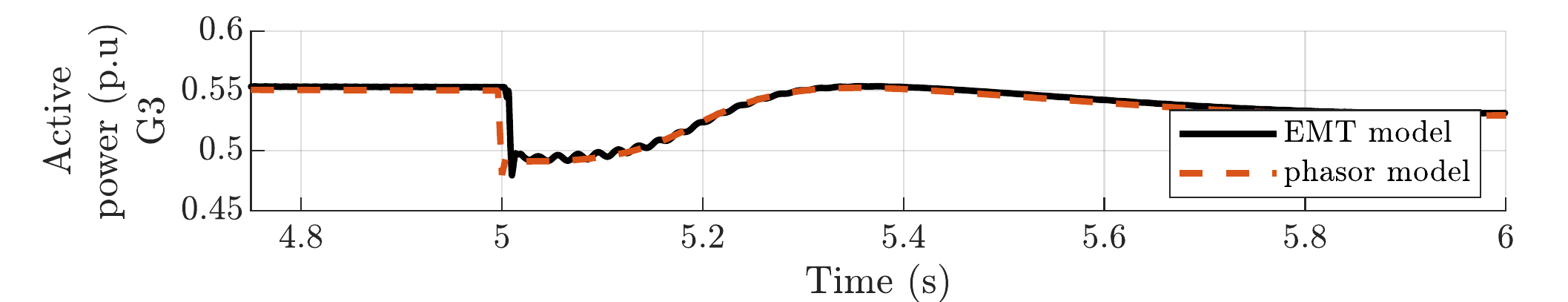}
 	\vspace{-3mm}
 	\caption{SG active power during a generator disconnection for each VSC model.}
 	\label{fig_vsc_disconectiongenerator_power_gen}
 \end{figure}

\begin{figure}[htb]
 	\vspace{-10pt}
 	\centering
 	\includegraphics[width=0.48\textwidth]{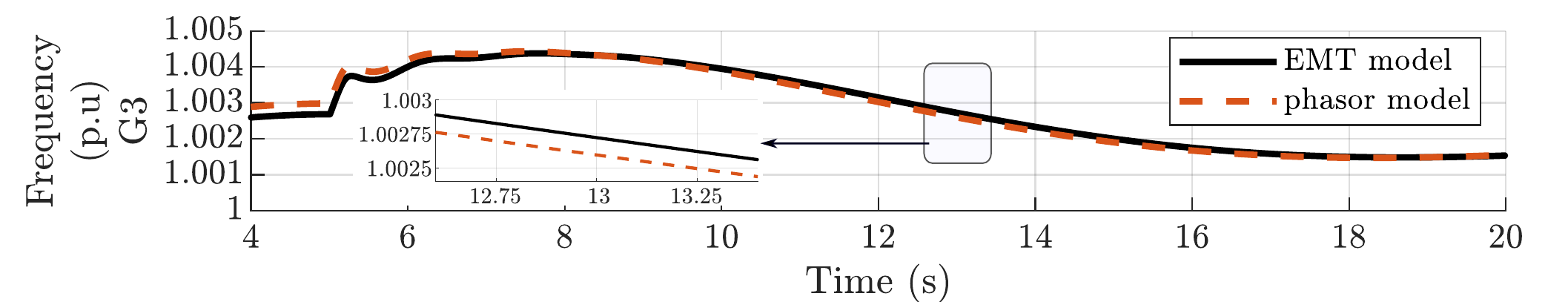}
 	\vspace{-3mm}
 	\caption{SG speed during a generator disconnection for each VSC model.}
 	\label{fig_vsc_disconectiongenerator_freq_gen}
 	\vspace{-1mm}
\end{figure}

\begin{figure}[htb!]
	\vspace{-10pt}
	\centering
	\includegraphics[width=0.48\textwidth]{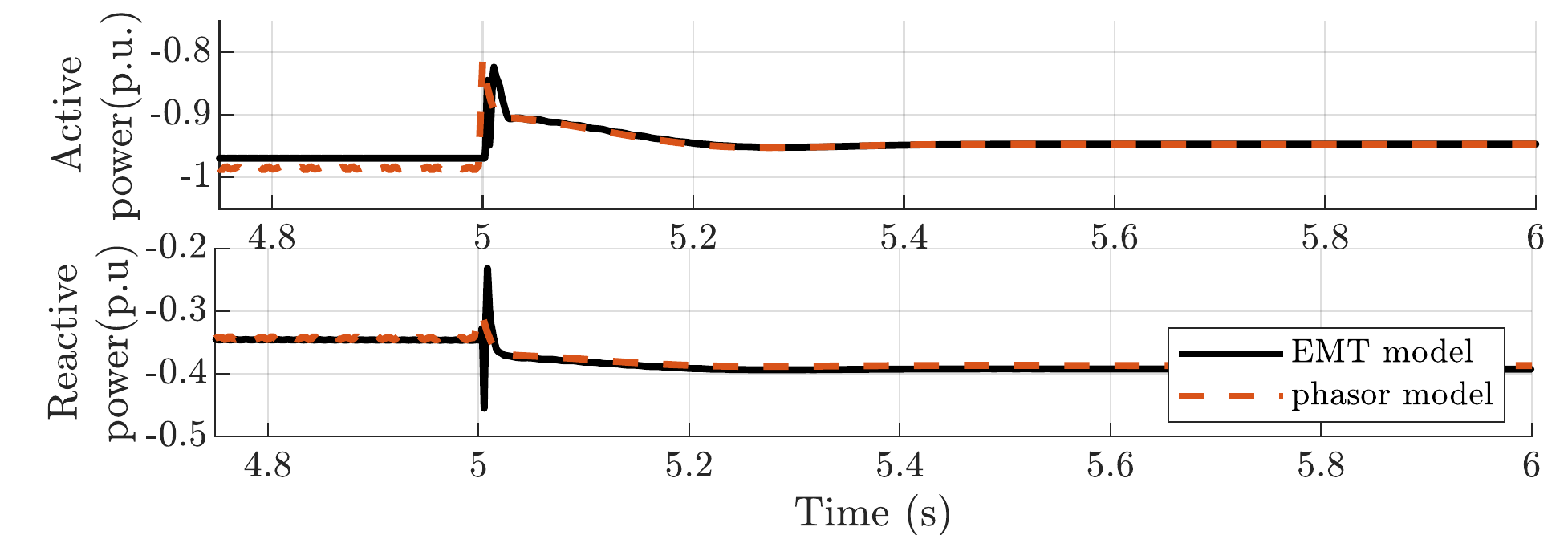}
	\vspace{-3mm}
	\caption{RES power during a generator disconnection for each VSC model.}
	\label{fig_vsc_disconectiongenerator_power_VSC}
	\vspace{-5mm}
\end{figure}

Figures \ref{fig_vsc_setpoint_power_VSC}-\ref{fig_vsc_disconectiongenerator_power_VSC} show the results for tests 1, 4 and 5 (setpoint tracking, asymmetric fault and loss of generation), respectively. 

In the setpoint tracking test, the VSC output power was nearly the same in both EMT and Phasor models.

During the asymmetric fault, SGs in EMT model showed an oscillatory torque produced by the transient, where only the average value is captured by the phasor model (Fig.~\ref{fig_vsc_asymetric_power_gen}). However, it should be highlighted that this effect is due to the simulation type (EMT or Phasor), and not due to the VSC model. Similar behaviour can be observed in the RES output power (Fig.~\ref{fig_vsc_asymetric_power_VSC}).

In the generator disconnection test, both EMT and phasor models presented similar responses, despite for an offset in the SGs speed (Fig. \ref{fig_vsc_disconectiongenerator_power_gen}-\ref{fig_vsc_disconectiongenerator_power_VSC}). This slight deviation in the phasor model might be due to model approximations as the stator fluxes derivatives are neglected.

The proper choice on using EMT or Phasor models will depend on the type of study and the system's size. Studies on small systems, focused on fast dynamics such as modulation and control studies, dynamic analysis of the PLL or detailed short circuit studies might require EMT simulation. On the other hand, studies on large systems, where electromechanical variables are being analysed, can be precisely simulated using phasor models in a fraction of the EMT simulation time. In the tests performed, the simulation with phasor models was 1277$\%$ faster in average when compared to simulations with EMT models.


\vspace{-3mm}
\subsection{Influence of RES}
\begin{figure}[htb]
	\vspace{-10pt}
	\centering
	\includegraphics[width=0.48\textwidth]{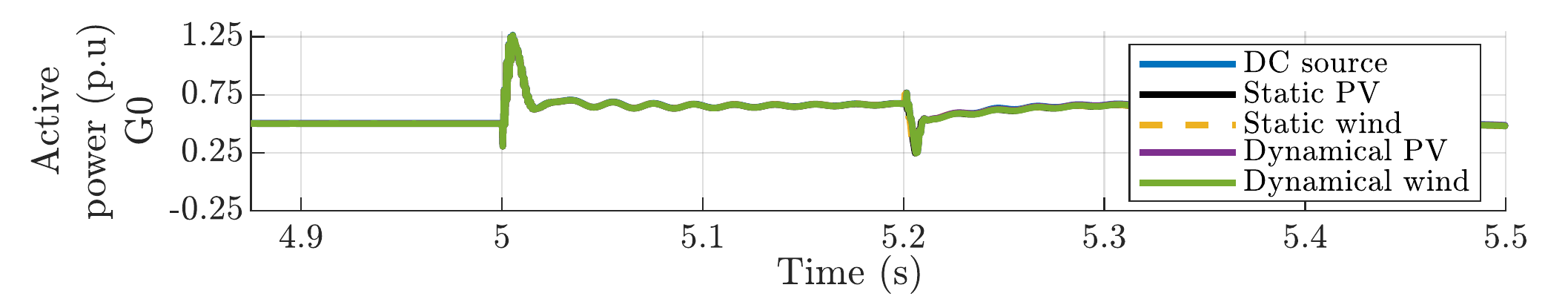}
	\vspace{-3mm}
	\caption{SG active power during a symmetric fault for each RES model.}
	\label{fig_RES_symmetric_power_gen}
	\vspace{-2mm}
\end{figure}

\begin{figure}[htb!]
	\vspace{-13pt}
	\centering
	\includegraphics[width=0.48\textwidth]{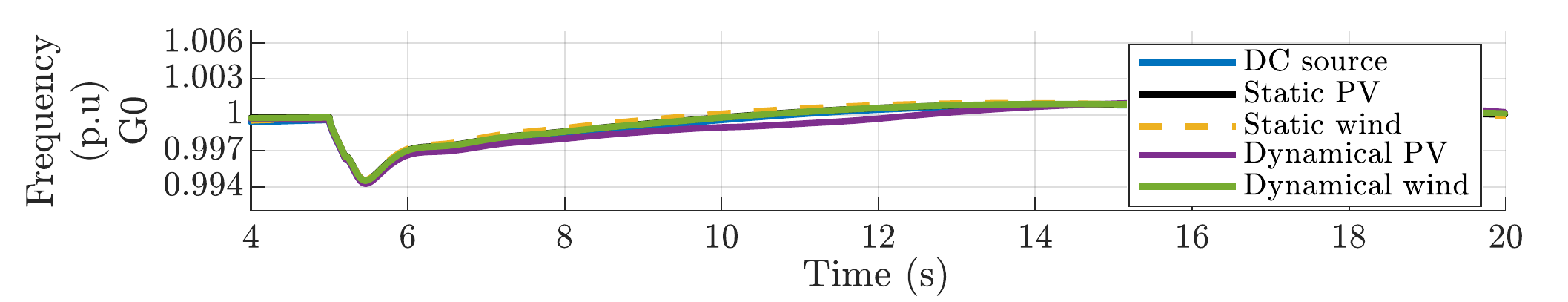}
	\vspace{-3mm}
	\caption{SG speed during a symmetric fault each RES model.}
	\label{fig_RES_symmetric_freq_gen}
	\vspace{-4mm}
\end{figure}

\begin{figure}[htb!]
	\vspace{-10pt}
	\centering
	\includegraphics[width=0.48\textwidth]{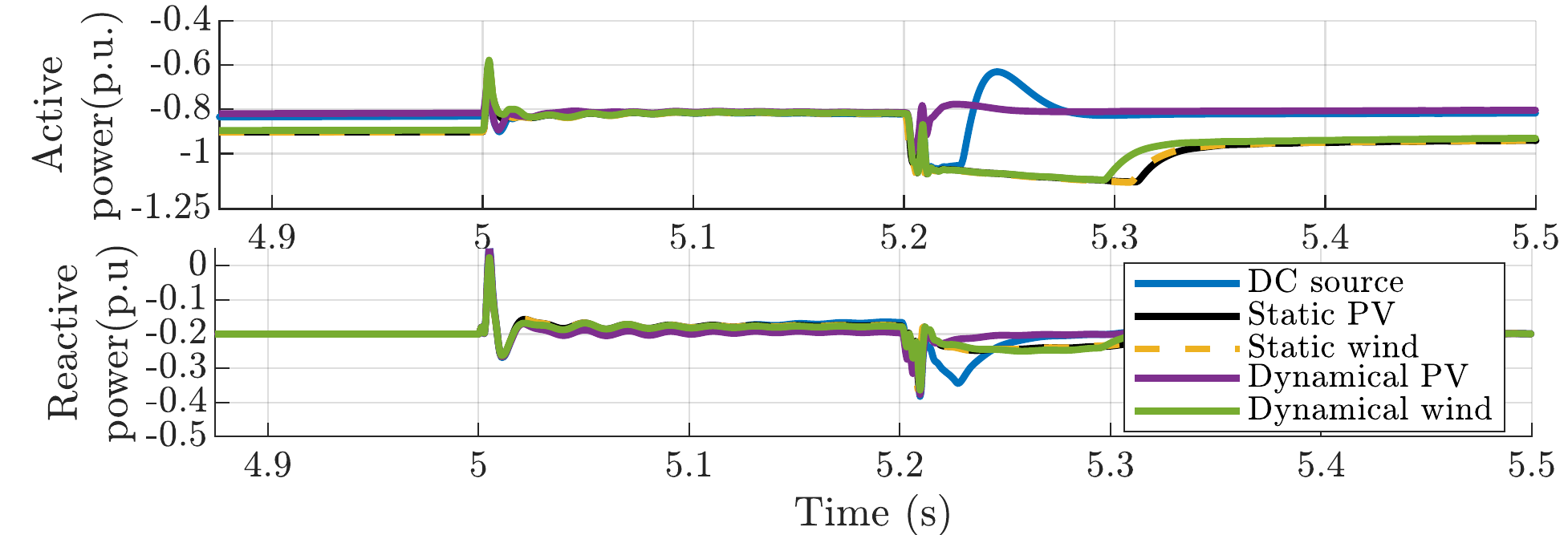}
	\vspace{-3mm}
	\caption{RES power during a symmetric fault for each RES model.}
	\label{fig_RES_symmetric_power_VSC}
	\vspace{-4mm}
\end{figure}

\begin{figure}[htb!]
	\vspace{-10pt}
	\centering
	\includegraphics[width=0.48\textwidth]{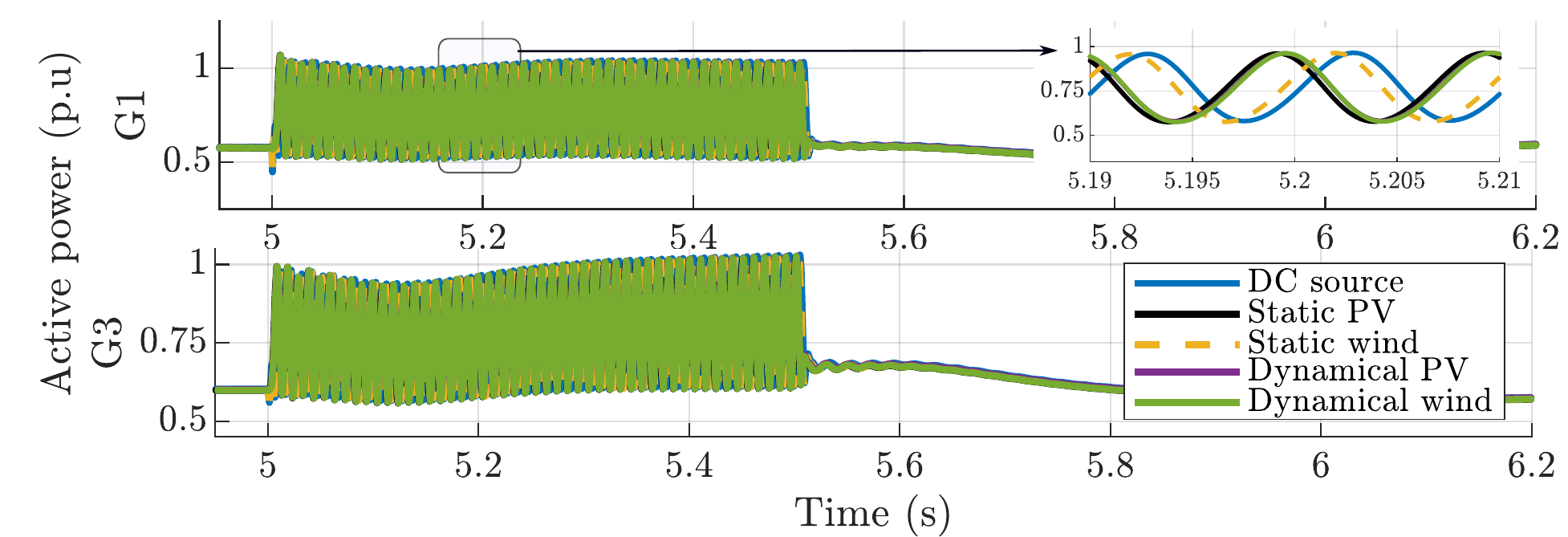}
	\vspace{-3mm}
	\caption{SGs active power during an asymmetric fault for each RES model.}
	\label{fig_RES_asymmetric_power_gen}
	\vspace{-6mm}
\end{figure}

\begin{figure}[htb!]
	
	\centering
	\includegraphics[width=0.48\textwidth]{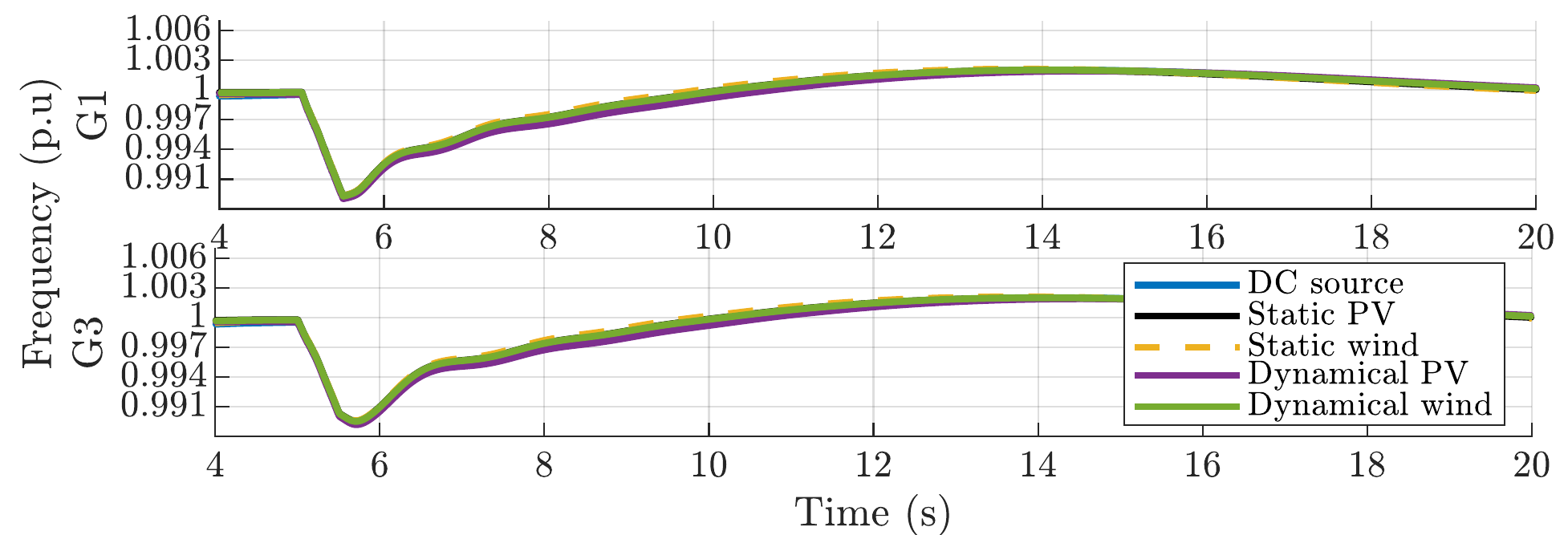}
	\vspace{-3mm}
	\caption{SGs speed during an asymmetric fault for each RES model.}
	\label{fig_RES_asymmetric_freq_gen}
	\vspace{-5mm}
\end{figure}

\begin{figure}[htb!]
	
	\centering
	\includegraphics[width=0.48\textwidth]{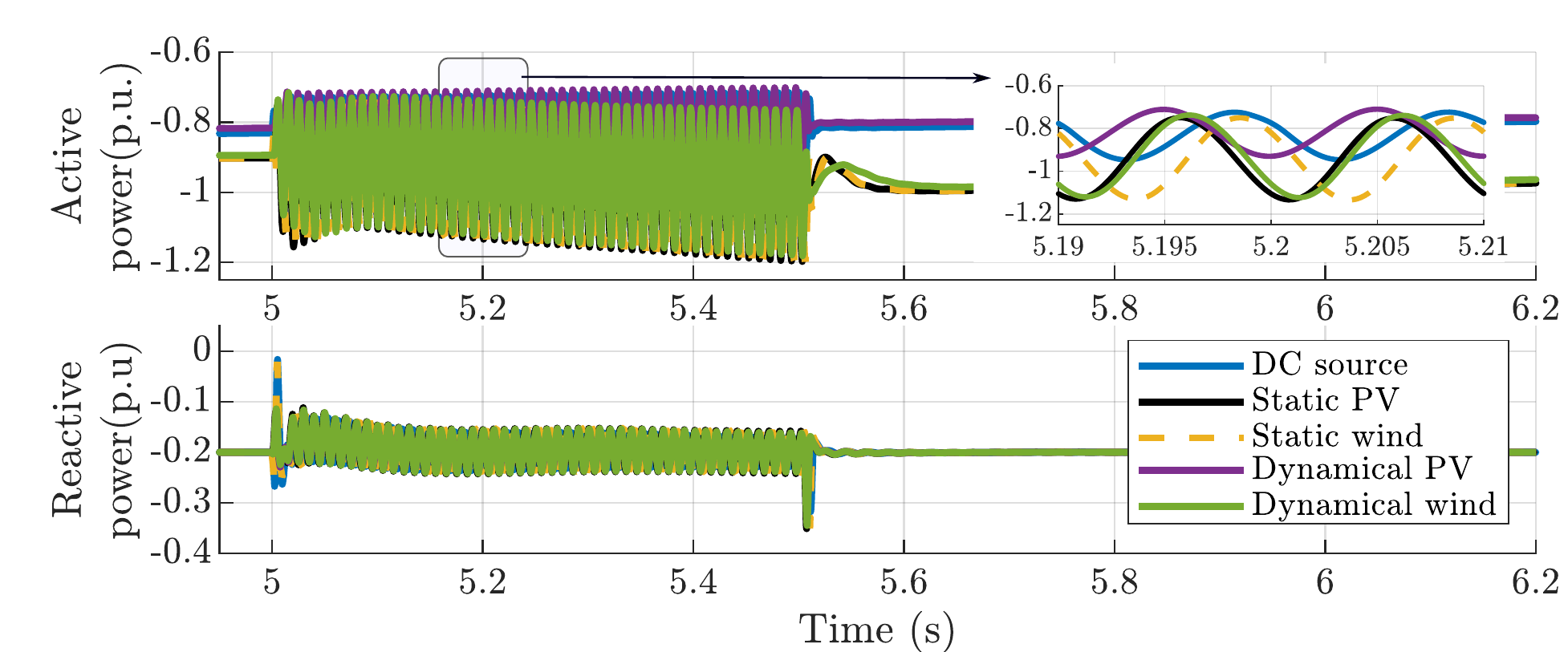}
	\vspace{-3mm}
	\caption{RES power during an asymmetric fault for each RES model.}
	\label{fig_RES_asymmetric_power_VSC}
	\vspace{-6mm}
\end{figure}

\vspace{-1mm}
Figures~\ref{fig_RES_symmetric_power_gen}-\ref{fig_RES_asymmetric_power_VSC} show the results for tests 3 and 4 (symmetric and asymmetric faults), respectively. 
By analysing the results, it can be observed that the RES model had minor influence on the SGs variables while the fault is active as the VSC's output power is governed by the fault ride through logic of the applied controller. 


 But in the VSC power, the difference is noticeable, especially post-fault. 
 After fault clearance, resynchronisation of the PLL yields a higher measured frequency than its nominal value, and consequently the frequency droop becomes active. This results in a reduced power demanded from the RES. At the same time, the terminal overvoltage leads to an increased emitted power to the grid (see Fig.~\ref{fig_RES_symmetric_power_VSC}). The dynamics associated with the DC link cause the delay in returning to the nominal power setpoint, especially visible for the static PV and both wind models. For the dynamical PV the P\&O algorithm prohibits the power variation of the source at the rates demanded by the frequency control loop and consequently the power imbalance in the DC link is less visible. Contrarily, DC source lacks a DC link and associated dynamics and therefore actuates directly according to the frequency droop signal.
 
 Therefore for studies in systems with low-share of RES and dominated by SGs, the ideal DC source would provide adequate accuracy. For studies with high-share of RES, the influence of the RES dynamics compared to static models needs to be further evaluated. 
 
 The fastest simulation was performed using the ideal DC source model, followed by static PV, which was 72,5$\%$ slower, followed by the static wind, which was 85,39 $\%$ slower than the ideal DC source. Finally, the slowest models were the dynamic PV (106.1$\%$ slower than the ideal DC source) and the dynamic wind  (164.82$\%$ slower than the ideal DC source). 
 


\FloatBarrier

\section{Conclusion}\label{sec_conclusions}
\vspace{-2mm}
This paper presented a comparative analysis amongst several models currently used to simulate power systems with RES in transient stability studies. Several models of SGs, transmission lines, converters and RES were compared in terms of precision and computational time. 

It could be observed that the most suitable models for the study of an electromechanical events in the system tested were the Model 2.2 \cite{IEEE_1110} without saturation, nominal  PI model, phasor model and ideal DC voltage source. These models were able to track the fundamental components of the variables been simulated and at the same time have less computational burden. However, if the system conditions or the study focus changes, other models might be more appropriate. For example, a system with high load might require modelling SG saturation to achieve a precise response. Moreover, if the converter switching is modelled, a frequency-dependent transmission line model would be needed to represent precisely the high-frequency components associated with the switching. The EMT model will be more adequate to represent fast oscillations during faults. In this study, the SGs dominate the system presented. However, if the share of RES increases, the RES model dynamics will have a more significant influence on the electromechanical events and thus dynamic RES models should be used.

The analyses provided aim to contribute to the definition of the most suitable models of key elements of modern power grids, where electromechanical and electromagnetic models are often simulated together.

\vspace{-1mm}
\bibliographystyle{IEEEtran}
\bibliography{ref.bib}

\end{document}